\documentclass[%
reprint,
groupedaddress,
amsmath,amssymb,
aps,
prl,
superscriptaddress]{revtex4-1}
\usepackage{graphicx}
\usepackage{dcolumn}
\usepackage{bm}
\usepackage{physics}
\usepackage{color}
\RequirePackage[colorlinks=true,linkcolor=blue,urlcolor=blue,hyperfootnotes=false,citecolor=blue]{hyperref}

\begin{document}
\title{Analog quantum control of magnonic cat states on-a-chip by a superconducting qubit}

\author{Marios Kounalakis}
 \email{marios.kounalakis@gmail.com}

\affiliation{Kavli Institute of Nanoscience, Delft University of Technology, 2628 CJ Delft, The Netherlands}

\author{Gerrit E. W. Bauer}
 \affiliation{WPI-AIMR, Tohoku University, 2-1-1, Katahira, Sendai 980-8577, Japan}
\affiliation{Kavli Institute of Nanoscience, Delft University of Technology, 2628 CJ Delft, The Netherlands}
\affiliation{Kavli Institute for Theoretical Sciences, University of the Chinese Academy of Sciences, 100190 Beijing, China}

\author{Yaroslav M. Blanter}
\affiliation{Kavli Institute of Nanoscience, Delft University of Technology, 2628 CJ Delft, The Netherlands}
\date{\today}

\begin{abstract}
We propose to directly and quantum-coherently couple a superconducting transmon qubit to magnons -- the quanta of the collective spin excitations, in a nearby magnetic particle.
The magnet’s stray field couples to the qubit via a superconducting quantum interference device (SQUID).
We predict a resonant magnon-qubit exchange and a nonlinear radiation-pressure interaction that are both stronger than dissipation rates and tunable by an external flux bias.
We additionally demonstrate a quantum control scheme that generates magnon-qubit entanglement and magnonic Schr{\"{o}}dinger cat states with high fidelity.
\end{abstract}
\maketitle

\emph{Introduction}.~Quantum magnonics is a rapidly growing field of research on magnetic devices operating in the quantum realm~\cite{chumak2022roadmap}.
Magnons, i.e., the quanta of the collective spin excitations in magnetic materials~\cite{chumak2015magnon,tabuchi2016quantum,lachance2019hybrid,rameshti2021cavity}, in the high-quality magnet yttrium iron garnet (YIG) couple strongly with microwave photons in superconducting resonators, i.e., with coupling rates far exceeding the system decay rates~\cite{huebl2013high,tabuchi2014hybridizing,zhang2014strongly}.
Quantum coherence of a macroscopic number of spins has been achieved by indirectly coupling a YIG sphere with a superconducting qubit in a microwave cavity~\cite{tabuchi2015coherent,lachance2017resolving,wolski2020dissipation,lachance2020entanglement}.
Magnon-photon coupling has also been demonstrated in optical setups, enabling microwave-optical transducers at the quantum level~\cite{viola2016coupled,zhang2016optomagnonic,osada2016cavity,haigh2016triple}.
These results establish magnons, along with phonons and photons, as promising carriers of quantum information in emerging hybrid quantum technologies~\cite{kurizki2015quantum,clerk2020hybrid,yuan2021quantum}.

An important milestone toward the realization of quantum devices is the ability to prepare and manipulate non-classical states of the many-spin system, that are strongly entangled, squeezed, or display negative Wigner functions~\cite{cahill1969density,kenfack2004negativity}, which may be expected in driven magnon-photon systems~\cite{elyasi2020resources}.
The macroscopic superpositions of coherent states or \textquotedblleft Schr{\"{o}}dinger cat\textquotedblright\ states~\cite{deleglise2008reconstruction} are especially attractive as essential building blocks for continuous-variable quantum information tasks, e.g., in fault-tolerant quantum computing~\cite{mirrahimi2014dynamically,Ofek2016,rosenblum2018fault,chamberland2020building}, quantum communication~\cite{chou2018deterministic,burkhart2021error}, and quantum simulation~\cite{flurin2017observing}, as well as for quantum metrology~\cite{zurek2001sub,munro2002weak,joo2011quantum} and fundamental studies of the quantum-to-classical transition~\cite{zurek2003decoherence,arndt2014testing}.
While such states can be prepared in superconducting cavities~\cite{Ofek2016,rosenblum2018fault,chamberland2020building,chou2018deterministic,burkhart2021error} and potentially in micromechanical resonators~\cite{asadian2014probing,khosla2018displacemon,kounalakis2019synthesizing,ma2021non}, magnetic systems offer a number of attractive and unique functionalities such as integration into spintronic circuits, unidirectional magnon propagation, as well as chiral couplings to phonons and photons~\cite{chen2019excitation,yu2020magnon,zhang2020unidirectional,bertelli2020magnetic}.

Recent theoretical proposals address the realization of such states in magnonic devices.
In the optical domain, magnonic cat states could be prepared by pulsed sideband driving~\cite{sun2021remote}, however this requires a much stronger coupling than appears possible to date.
Creating such states in an ellipsoid ferromagnet inside a microwave cavity might be possible, but is hindered by the requirement of collapsing the cavity field in a high-photon-number state~\cite{sharma2021cat}.
Quantum states of a driven ferromagnet can be generated in a cavity when resonantly coupled to a qubit~\cite{sharma2022protocol} following protocols from cavity and circuit quantum electrodynamics~\cite{law1996arbitrary,hofheinz2009synthesizing} based on the \emph{digital quantum computing} paradigm of sequential qubit operations (quantum gates)~\cite{deutsch1989quantum,dodd2002universal}.
However, cumulative gate errors as well as long gate operation times hinder the preparation of macroscopic quantum superpositions.
In addition, despite their flexibility, digital schemes are very demanding on resources, compared to more robust analog approaches~\cite{para2020digital}.
It is therefore desirable to develop {\em analog} schemes to prepare and control quantum states in which the system evolves under natural interactions without clocked external operations.

Here we propose a hybrid device comprising a YIG particle directly coupled to a planar superconducting qubit via magnetic stray fields, which can additionally be used to synthesize high-fidelity magnonic cat states by analog means.
The direct magnon-qubit flux-mediated interaction is sufficiently strong at practical distances such that a mediating microwave cavity is not needed.
Moreover, the interaction can be tuned in-situ, providing a lot of flexibility in constructing artificial magnonic quantum networks~\cite{rusconi2019hybrid}.
Specifically, the system features two distinct and tunable couplings.
First, we find a resonant magnon-qubit exchange interaction, akin to the Jaynes-Cummings model~\cite{jaynes1963comparison,tabuchi2015coherent}, which can be used to create quantum states by digital protocols~\cite{law1996arbitrary,sharma2022protocol}.
Second, we find a strong and nonlinear interaction analogous to the radiation-pressure in optomechanics~\cite{shevchuk2017strong,rodrigues2019coupling,kounalakis2020flux}, which has never been explored in magnon-qubit systems before.
We demonstrate that this coupling can be resonantly enhanced by dynamically driving the qubit to generate robust quantum superpositions of coherent states in a purely analog fashion, with a fidelity that is limited only by the magnon lifetime.

\begin{figure}[t]
  \begin{center}
  \includegraphics[width=1.05\linewidth]{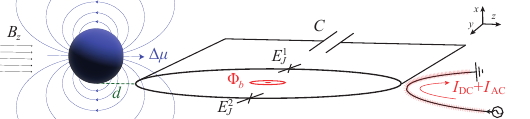}
  \end{center}
  \caption{
  {Proposed circuit architecture.}
  A YIG particle with uniform magnetization oriented by an in-plane field $B_z$ close to a flux-tunable transmon qubit, formed by a SQUID loop and a capacitor $C$ in parallel.
  The magnetic fluctuations $\Delta\mu$ induce a flux in the SQUID that modulates its inductive energy and therefore the qubit frequency.
  An external flux bias $\Phi_b$ can be applied locally via control lines carrying DC and AC currents.
  }
  \label{fig:scheme}
\end{figure}

\emph{The device}.~The superconducting element in our circuit is a flux-tunable transmon qubit~\cite{koch2007charge}, i.e., a superconducting quantum interference device (SQUID) shunted by a capacitor $C$, as depicted in Fig.~\ref{fig:scheme}.
The former is a superconducting loop interrupted by two Josephson junctions, with Josephson energies $E_{J}^{1}$ and $E_{J}^{2}$.
Its inductive energy depends nonlinearly on the superconducting phase difference $\hat{\delta}$ and an externally applied flux $\Phi_{b}$~\cite{zorin1996quantum},
\begin{equation}
\mathcal{E}_{\mathrm{ind}}(\hat{\delta},\phi_{b})=-E_{J}^{\mathrm{max}}S(\phi_{b})\cos{\left(  \hat{\delta}-\arctan{(a_{J}\tan{\phi_{b}})}\right)  },\label{eq:Hind_transmon}
\end{equation}
where $\phi_{b}\dot{=}\pi\Phi_{b}/\Phi_{0},$ ${S(\phi_{b})=\sqrt{\cos^{2}{\phi_{b}}+a_{J}^{2}\sin^{2}{\phi_{b}}}},$ and the imbalance between the Josephson energies or \emph{SQUID asymmetry} $a_{J}=\vert E_{J}^{1}-E_{J}^{2}\vert /E_{J}^{\mathrm{max}}$ is an important design parameter (see below), with $E_{J}^{\mathrm{max}}\dot{=}E_{J}^{1}+E_{J}^{2}$.

The transmon is described by the Hamiltonian ${\hat{H}_{0}^{T}=4E_{C}\hat{N}^{2}+\mathcal{E}_{\mathrm{ind}}(\hat{\delta},\Phi_{b})}$, where $E_{C}=e^2/(2C)$ is the charging energy and $\hat{N}$ (operator conjugate to $\hat{\delta}$) represents the number of tunneling Cooper pairs~\cite{vool2017introduction}.
The operators $\hat{N},~\hat{{\delta}}$ can be expressed in terms of bosonic annihilation (creation) operators $\hat{c}^{(\dagger)}$~\cite{vool2017introduction},
\begin{equation}
\hat{N}=iN_\mathrm{zpf}(\hat{c}^{\dagger}-\hat{c}),~\hat{{\delta}}=\delta_\mathrm{zpf}(\hat{c}+\hat{c}^{\dagger}),
\label{eq:TransmonOperators}
\end{equation}
where $N_\mathrm{zpf}= \left[   E_{J}^{\mathrm{max}} S(\phi_{b}) / ( 32E_C ) \right]^{1/4}$ and $\delta_\mathrm{zpf}= \left(  2E_{C}/ \left[E_{J}^{\mathrm{max}}S(\phi_{b})\right] \right) ^{1/4}$.
In the transmon regime ($E_{J}^{\mathrm{max}}S(\phi_{b})\gg E_{C}$), $\hat{H}_{0}^{T}=\hbar\omega_{q}\hat{c}^{\dagger}\hat{c}-\frac{E_{C}}{2}\hat{c}^{\dagger}\hat{c}^{\dagger}\hat{c}\hat{c}$, where $\hbar\omega_{q}={\sqrt{8E_{C}E_{J}^{\mathrm{max}}S(\phi_{b})}-E_{C}}$ is the transmon excitation energy and the second (self-Kerr) nonlinear term defines the anharmonicity ${-E_{C}}$~\cite{koch2007charge}.

The magnet is a YIG particle, without loss of generality chosen here to be a sphere with radius $R_{\mathrm{YIG}}$, placed at an in-plane center-to-center distance $d+R_\mathrm{SQUID}$ from the SQUID, as depicted in Fig.~\ref{fig:scheme}.
An in-plane magnetic field $B_{z}\mathbf{\hat{z}}$ orients the magnetization $\mathbf{M}$ or spin angular momentum $\mathbf{S}=-\frac{4}{3}\pi R_{\mathrm{YIG}}^{3}\mathbf{M/}\gamma_{0}$, where $\gamma_{0}$ is the modulus of the gyromagnetic ratio~\cite{rameshti2021cavity}.
The fundamental excitation is a uniform precession or \emph{Kittel} mode with ferromagnetic resonance (FMR) frequency $\omega_{m}=\gamma_{0}(B_{z}+B_\mathrm{ani})$, where $B_\mathrm{ani}$ is the anisotropy field~\cite{stancil2009spin}.

The Hamiltonian of the magnetic order can be mapped on a quantum harmonic oscillator by the leading term of the Holstein-Primakoff expansion in terms of bosonic operators $m^{(\dagger)}$ that create (annihilate) a magnon~\cite{rameshti2021cavity}.
Omitting the zero-point energy $\hbar\omega_{m}/2$, a weakly excited Kittel mode is well-described by $\hat{H}_{0}^{M}=\hbar\omega_{m}\hat{m}^{\dagger}\hat{m}$~\cite{rameshti2021cavity}.
The amplitudes of the magnetic excitations are $\Delta\hat{\mu}_{x}=\mu_{\mathrm{zpf}}(\hat{m}+\hat{m}^{\dagger}),$ $\Delta\hat{\mu}_{y}=i\mu_{\mathrm{zpf}}(\hat{m}-\hat{m}^{\dagger}),$ and $\Delta\hat{\mu}_{z}=\hbar\gamma_{0}\hat{m}^{\dagger}\hat{m},$ where $\mu_{\mathrm{zpf}}=\hbar\gamma_{0}\sqrt{N_S/2}$, and $N_S$ is the total number of spins.

The magnetic moment emits a stray field $\mathbf{{B_{\mathrm{YIG}}}}$ that induces a flux through the SQUID loop, thereby modulating its inductive energy and the qubit frequency.
According to Eq.~(\ref{eq:Hind_transmon}) by $\phi_{b}\rightarrow\phi_{b}+\phi\left(  \Delta\bm{\hat{\mu}}\right)$ [see Supplementary Information (SI)~\cite{SI_transmonmagnon} for the derivation],
\begin{align}
\mathcal{E}_{\mathrm{ind}}(\hat{\delta},\phi_{b},\Delta\bm{\hat{\mu}}) &=-sE_{J}^{\mathrm{max}}\left\{  \left[  \cos{\phi_{b}}-\phi(\Delta\bm{\hat{\mu}})\sin{\phi_{b}}\right]  \cos{\hat{\delta}}\right.  \nonumber\\
&  \left.  +a_{J}\left[  \sin{\phi_{b}}+\phi(\Delta\bm{\hat{\mu}})\cos{\phi_{b}}\right]  \sin{\hat{\delta}}\right\}  ,
\end{align}
where $s\dot{=}\mathrm{sgn}[\cos{\phi_{b}}]$ for $\phi(\Delta\bm{\hat{\mu}})\ll1$, noting that ${\phi(\Delta\bm{\hat{\mu}})\lesssim10^{-3}}$ for typical parameters.
We thus arrive at a total Hamiltonian, $\hat{H}=\hat{H}_{0}^{M}+\hat{H}_{0}^{T}+\hat{H}_{\mathrm{int}}$ with
\begin{equation}
\hat{H}_{\mathrm{int}}=\frac{E_{J}^{\mathrm{max}}\phi(\Delta\bm{\hat{\mu}})}{S(\phi_{b})}\left[  \frac{\sin{2\phi_{b}}}{2}\left(1-a_{J}^{2}\right)  \cos{\hat{\tilde{\delta}}}-a_{J}\sin{\hat{\tilde{\delta}}}\right],
\label{eq:InteractionHamiltonian}
\end{equation}
where $\hat{\tilde{\delta}}\dot{=}\hat{\delta}-\arctan{(a_{J}\tan}${{$\phi_{b}$}}${)}$.

In the far-field limit the induced flux reads
\begin{equation}
{\phi(\Delta\bm{\hat{\mu}})=\iint{\mathbf{{B_{\mathrm{YIG}}}}}\left(\Delta{\mathbf{\bm{\hat{\mu}}}}\right)  {d\mathbf{{A}}}} =\frac{\mu_{0}}{4\Phi_{0}d_{\mathrm{min}}}\sum_{i=x,y,z}{I_{i}\Delta\hat{\mu}_{i}},
\end{equation}
where $A$ is the loop area, $d_{\mathrm{min}}=\sqrt{R_{\mathrm{YIG}}^2+d^2}$ and $I_{x},I_{y},I_{z}=\mathcal{O}\left(  1\right)$ are dimensionless geometrical factors.
For a point magnetic dipole, in the limit of a large loop radius, and assuming $d=R_{\mathrm{YIG}}$ (${d_{\mathrm{min}}=\sqrt{2}R_{\mathrm{YIG}}}$) we have $I_{y}=0$, ${I_{x}\simeq-1}$~\cite{rusconi2019hybrid}, therefore,
\begin{equation}
\phi(\Delta\bm{\hat{\mu}})=-\frac{\mu_{0}\mu_{\mathrm{zpf}}}{4\Phi_{0}d_{\mathrm{min}}}(\hat{m}+\hat{m}^{\dagger}).
\label{eq:PhiDeltaMu}
\end{equation}
Note that we disregarded the magnon number fluctuations since $\hbar\gamma_0\ll\mu_{\mathrm{zpf}}$ for \(N_S \gg 1\).

Expanding to order ${\tilde{\delta}}^{2}$ in Eq.~(\ref{eq:InteractionHamiltonian}) and using Equations~(\ref{eq:TransmonOperators}) and (\ref{eq:PhiDeltaMu}), the interaction reads
\begin{equation}
\hat{H}_{\mathrm{int}}/\hbar=J(\hat{c}^{\dagger}\hat{m}+\hat{c}\hat{m}^{\dagger})+g_{\mathrm{rp}}\hat{c}^{\dagger}\hat{c}(\hat{m}+\hat{m}^{\dagger}),
\label{eq:Hint_quantum}
\end{equation}
where we disregarded fast-rotating terms $(\hat{c}^{\dagger})^{n}\hat{m}^{(\dagger)}$ (rotating-wave approximation) because ${J,g_{\mathrm{rp}}\ll\omega_{q,m}}$.
Terms of $\mathcal{O}[\hat{\tilde{\delta}}^4]$ in Eq.~(\ref{eq:InteractionHamiltonian}) lead to interactions that are irrelevant in the qubit manifold, but cause small corrections to $g_\text{rp}$ and $J$~\cite{SI_transmonmagnon}.

\emph{Magnon-qubit couplings}.~The first term in Eq.~(\ref{eq:Hint_quantum}),
\begin{equation}
J=\frac{\mu_{0}\mu_{\mathrm{zpf}}}{4d_{\mathrm{min}}\Phi_{0}}\frac{a_{J}\left(  2E_{C}(E_{J}^{\mathrm{max}})^{3}\right)  ^{1/4}}{\hbar\lbrack S(\phi_{b})]^{5/4}} ,
\label{eq:Jcoupling}
\end{equation}
describes the coherent exchange between qubit and magnon excitations, similar to the Jaynes-Cummings model of light-matter interaction~\cite{jaynes1963comparison} and the effective magnon-qubit coupling in cavities~\cite{tabuchi2015coherent}.
The nonlinear term in Eq.~(\ref{eq:Hint_quantum}) has a coupling strength
\begin{equation}
g_{\mathrm{rp}}=\frac{\mu_{0}\mu_{\mathrm{zpf}}}{16d_{\mathrm{min}}\Phi_{0}}\frac{\sqrt{8E_{J}^{\mathrm{max}}E_{C}}(1-a_{J}^{2})\sin{2\phi_{b}}}{\hbar\lbrack S(\phi_{b})]^{3/2}} ,
\label{eq:g_rp}
\end{equation}
similar to the radiation pressure in optomechanical systems~\cite{aspelmeyer2014cavity} or the optical photon-magnon coupling~\cite{viola2016coupled}.

\begin{figure}[t]
  \begin{center}
  \includegraphics[width=\linewidth]{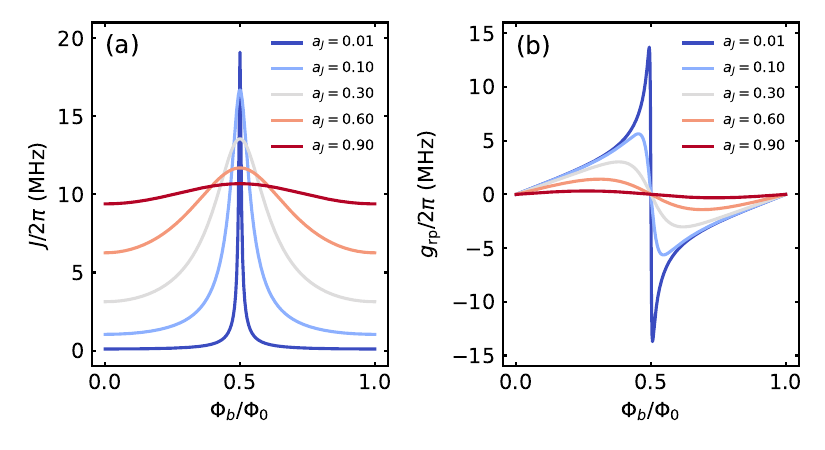}
  \end{center}
  \caption{
  {Couplings vs flux bias}. 
  Coupling strength of the magnon-qubit exchange interaction $J$~(a) and ``photon-pressure" interaction $g_{\mathrm{rp}}$~(b) as a function of the applied flux bias $\Phi_b$, for different values of the SQUID asymmetry $\alpha_J$.
  }
  \label{fig:couplings}
\end{figure}

In Fig.~\ref{fig:couplings} we plot both coupling strengths as a function of the applied flux $\Phi_{b}$ and SQUID asymmetry $a_{J}$, including higher-order corrections~\cite{SI_transmonmagnon}.
We assume typical transmon parameters $E_{J}^{\text{max}}/h=50$~GHz, $E_{C}/h=200$~MHz and YIG radius $R_{\mathrm{YIG}}=3~\mathrm{\mu m}$ leading to $N_S=2.4\cross10^{12}$ for typical densities~\cite{tabuchi2015coherent}.
$J$ has a maximum at $\phi_{b}=\pi/2$, vanishes for a fully symmetric SQUID ($a_{J}=0$) and approaches a constant when highly asymmetric.
While here we assumed that the junction capacitances are the same, in the SI we treat the more general case including a finite capacitance asymmetry and find that $J=0$ when one junction vanishes ($a_{J}=1$), as expected~\cite{SI_transmonmagnon}.
On the other hand, $g_{\mathrm{rp}}$ changes sign at $\phi_{b}=\pi/2$ and vanishes for $a_{J}\rightarrow1$.
Both its maximal value and the optimal bias point depend on $a_{J}$, which is fixed by sample design and fabrication~\cite{kounalakis2019nonlinear}.

Quantum manipulations require operation in the strong coupling regime ${J,g_{\mathrm{rp}}>\kappa,1/T_{1,2}}$, where ${\kappa=\omega_{m}\alpha_{G}}$ is the magnon decay rate in terms of the Gilbert damping constant $\alpha_{G}$, while $T_{1}$ and $T_{2}$ are the transmon relaxation and dephasing times, respectively.
As shown in Fig.~\ref{fig:couplings}(a), $J$ is minimal at $\phi_{b}=k\pi~(k\epsilon\mathbb{Z})$ where the qubit is insensitive to flux noise (see Fig.~\ref{fig:FreqVflux}~\cite{SI_transmonmagnon}).
However, highly asymmetric transmons operate equally well around the stationary points $\phi_{b}=(2k+1)\pi/2$, at the cost of a narrower tuning range~\cite{hutchings2017tunable}.
The asymmetry parameter strikes a compromise between strong coupling, acceptable tuning range, and high coherence.
For example, choosing $a_{J}=0.6$ at $\phi_{b}=\pi/2 ,$ we have $J/(2\pi)\gtrsim10$~MHz, therefore the system is in the strong coupling regime with $T_{1,2}\simeq20\,\mathrm{\mu s}$~\cite{langford2017experimentally} and typical Gilbert damping parameters $\alpha_{G}\sim10^{-5}-10^{-4}$~\cite{tabuchi2015coherent,klingler2017gilbert,rameshti2021cavity}.
The interaction can be switched on and off easily by fast flux-pulses shifting the qubit frequency~\cite{langford2017experimentally} and algorithmic sequences of qubit control pulses may create arbitrary quantum states of magnons~\cite{law1996arbitrary,hofheinz2009synthesizing,sharma2022protocol}.
However, multiple qubit pulses propagate errors that can prevent the generation of a given state.
Moreover, the finite pulse duration, typically around $10-100$~ns~\cite{kjaergaard2020superconducting}, limits the number of operations that can be performed within the relaxation times.
Given the relatively short magnon lifetime at GHz frequencies, ranging from hundreds of nanoseconds to a few microseconds~\cite{tabuchi2015coherent,klingler2017gilbert,rameshti2021cavity}, such digital schemes appear less attractive.

Rather than building magnon superpositions pulse-by-pulse, large cat states can be generated by the nonlinear radiation pressure, which couples magnonic displacements with qubit frequency shifts.
This coupling, in the interaction picture $g_{\mathrm{rp}}\hat{c}^{\dagger}\hat{c}(\hat{m}e^{-i\omega_mt}+\hat{m}^{\dagger}e^{i\omega_mt})$, can be activated in the \emph{ultra-strong coupling} regime ${g_{\mathrm{rp}}\gtrsim\omega_m}$~\cite{khosla2018displacemon,kounalakis2020flux}, or by applying a stroboscopic qubit pulse train with period $T=\pi/\omega_m$ and pulse duration ${\ll T}$~\cite{tian2005entanglement}.
Alternatively, it may be activated in a tripartite configuration with an additional qubit, enabling arbitrary magnonic states via pulsed schemes~\cite{kounalakis2019synthesizing}.
However, these approaches encounter the difficulty that magnon frequencies are larger than ${100~\mathrm{MHz}\gg g_{\mathrm{rp}}},$ while qubit gates below $10$~ns lead to interference of higher transmon levels~\cite{kjaergaard2020superconducting}.
Here we propose to enhance the nonlinearity by a time-dependent modulation of the coupling as proposed for trapped ions and mechanical systems~\cite{kielpinski2012quantum,liao2014modulated} based on a parametric flux-modulation of the SQUID loop, which can be implemented by local control striplines~\cite{mckay2016universal}.

\emph{Cat state preparation protocol}.~The coupling Eq.~(\ref{eq:g_rp}) reduces for a symmetric SQUID to ${g_{\mathrm{rp}} =\frac{\omega_{p}\sin{\phi_{b}}}{\sqrt{\left\vert \cos{\phi_{b}}\right\vert }} \frac{\mu_{0}\mu_{\mathrm{zpf}}}{8\Phi_{0}d_\mathrm{min}}  }\rightarrow \omega_{p}\frac{\mu_{0}\mu_{\mathrm{zpf}}\phi_{\mathrm{ac}}} {8\Phi_{0}d_\mathrm{min}}
\cos{(\omega_{\mathrm{ac}}t)}$, where in the second step we used a weak ac bias ${\phi_{b}=\phi_{\mathrm{ac}}\cos{(\omega_{\mathrm{ac}}t)}}$ (${\phi_{\mathrm{ac}}\ll1}$) and defined the Josephson plasma frequency ${\omega_{p}=\sqrt{8E_{J}^{\mathrm{max}}E_{C}}/\hbar}$.
The Hamiltonian, in the rotating frame $\hat{U}(t)=e^{i{\omega_{\mathrm{ac}}}t\hat{m}^{\dagger}\hat{m}}$, reads
\begin{equation}
\hat{\widetilde{H}} =\hat{U}^{\dagger}\hat{H}\hat{U}=\hat{H}_{0}^{T}+\hbar\delta\hat{m}^{\dagger}\hat{m}+\hbar\tilde{g}_{\mathrm{rp}}\hat{c}^{\dagger}c(\hat{m}+\hat{m}^{\dagger}),
\label{eq:H_system}
\end{equation}
with ${\delta=(\omega_{m}-{\omega_{\mathrm{ac}}}})$ and $\tilde{g}_{\mathrm{rp}}=  \frac{\mu_{0}\mu_{\mathrm{zpf}}\phi_{\mathrm{ac}}}{16\Phi_{0}d_\mathrm{min}} \omega_{p}$.
Here we employed the rotating wave approximation by omitting rapidly varying interaction terms $\tilde{g}_{\mathrm{rp}}\hat{c}^{\dagger}c\hat{m}^{(\dagger)}e^{\pm i2\omega_{\mathrm{ac}}t}$ $(2{\omega_{ac}}\gg\tilde{g}_{\mathrm{rp}})$ and assumed unperturbed qubit eigenstates since $\omega_{\mathrm{ac}}\ll\omega_{q}$.
Flux modulation is experimentally well established~\cite{mckay2016universal} without requiring additional circuitry, in contrast with modulating $E_C$, $d_\mathrm{min}$ or $\mu_{\mathrm{zpf}}$.
In the configuration of Fig.~\ref{fig:scheme} the flux bias line and the magnet are spatially separated to minimize crosstalk.
The remaining weak microwave field results in a slight coherent displacement of the magnon state, which does not perturb the interaction dynamics.

The resonant modulation $({\delta=0})$ generates coherent magnon displacements conditioned by the qubit, leading to macroscopic quantum superpositions of magnetization by the following protocol.
Starting with both systems in their ground state and applying a $R_{\hat{y},\frac{\pi}{2}}$ qubit pulse~\cite{vlastakis2013deterministically,langford2017experimentally}, creates the superposition state ${\vert +\rangle_q=(\vert 0\rangle_q+\vert 1\rangle_q)/\sqrt{2}}$.
After turning on the modulation, the system evolves (according to $\hat{\widetilde{H}}$) into ${(\vert 0\rangle_q\vert 0\rangle_m+e^{i\theta(t)}\vert 1\rangle_q\vert \beta(t)\rangle_m)/\sqrt{2}}$, where ${\beta(t)=(\tilde{g}_{\mathrm{rp}}/\delta)\left(e^{-i\delta t}-1\right)},~{\theta(t)=(\tilde{g}_{\mathrm{rp}}/\delta)^2(\delta t-\sin{\delta t})}$~\cite{asadian2014probing}.
Switching off the flux modulation at $t=\tau$ leaves the system in a highly-entangled hybrid Bell-cat state~\cite{vlastakis2015characterizing} $\frac{1}{2}\left[\vert +\rangle_q(\vert 0\rangle+e^{i\theta(\tau)}\vert \beta(\tau)\rangle)_m+\vert -\rangle_q(\vert 0\rangle-e^{i\theta(\tau)}\vert \beta(\tau)\rangle)_m\right]$.

The protocol is concluded by applying a second $R_{\hat{y},\frac{\pi}{2}}$ rotation followed by a strong projective measurement that collapses the qubit state.
If the measurement yields $\vert 0\rangle_q$ ($\vert 1\rangle_q$) the magnet is left in a macroscopic superposition of coherent states, i.e., an \emph{even (odd) cat state} $\vert \psi\rangle_{_\mathrm{odd}^\mathrm{even}}=(\vert 0\rangle_m\pm \vert \beta(\tau)\rangle_m)/\mathcal{N}$, where $\mathcal{N}=\sqrt{2}(1\pm e^{-\left\vert\beta(t)\right\vert^2/2})^{1/2}$, and ${\beta(\tau)=-i\tilde{g}_{\mathrm{rp}}\tau}$ for ${\delta\rightarrow0}$.
Following a coherent magnon displacement of amplitude ${-\beta(\tau)/2}$ this state is equivalent to ${(\vert {-\beta(\tau)/2}\rangle_m\pm \vert \beta(\tau)/2\rangle_m)/\mathcal{N}}$, consisting of only even (odd) magnon number states.

We can model these operations by the quantum statistical Lindblad master equation~\cite{johansson2012qutip}
$\dot{\rho}=\frac{i}{\hbar}[\rho,\hat{\widetilde{H}}]+\omega_{m}\alpha_{G}\left(n_\mathrm{th}\mathcal{L}[\hat{m}^\dagger]\rho+(n_\mathrm{th}+1)\mathcal{L}[\hat{m}]\rho\right)+\frac{1}{T_1}\mathcal{L}[\hat{c}]\rho+\frac{1}{T_2}\mathcal{L}[\hat{c}^\dagger\hat{c}]\rho,$ where the superoperator $\mathcal{L}[\hat{o}]\rho=(2\hat{o}\rho\hat{o}^\dagger-\hat{o}^\dagger\hat{o}\rho-\rho\hat{o}^\dagger\hat{o})/2$ describes the bare dissipation channels, and ${n_\mathrm{th}=1/[\exp(\hbar\omega_m/(k_BT))-1]}$ is the number of thermally excited magnons at temperature $T$.
YIG's weak magnetic anisotropy ${B_\mathrm{ani}=-2.5}$~mT~\cite{klingler2017gilbert} enables working at sub-GHz frequencies with low magnon decay rates $\omega_{m}\alpha_{G}$ and weak magnetic fields.
We chose ${\omega_m/(2\pi)=500}$~MHz and ${T=5}$~mK (for higher temperatures see~\cite{SI_transmonmagnon}) and typical qubit relaxation and dephasing times $T_1=T_2=20~\mathrm{\mu s}$~\cite{kjaergaard2020superconducting}.
The required in-plane magnetic field ${\sim20}$~mT does not compromise the qubit performance~\cite{krause2021magnetic}.
We model the transmon $\hat{H}_{0}^{T}$ as a three-level system with anharmonicity ${-E_C}$, although there is no leakage from the qubit subspace during the protocol, and solve the dynamics numerically in a basis of up to 140 magnon levels.

\begin{figure}[t]
  \begin{center}
  \includegraphics[width=\linewidth]{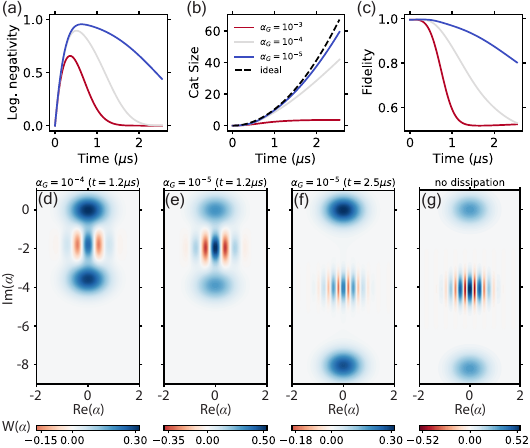}
\end{center}
  \caption{
  Magnonic cat states.
  (a)~Logarithmic negativity $E_N$ showcasing the evolution of entanglement in the bipartite system for ${10^{-5}\le\alpha_{G}\le10^{-3}}$.
  (b)~Corresponding magnon cat size $S$, where the dashed curve plots the ideal case $S=\left\vert\tilde{g}_{\mathrm{rp}}t\right\vert^2$.
  (c)~Fidelity of the prepared state to target $\vert\psi\rangle_{\mathrm{even}}$.
  Bottom row:~Wigner function of prepared magnonic cats with fidelities \(\mathcal{F} \ge 80~\%\) for $\alpha_G=10^{-4}$(d), $10^{-5}$(e-f) and no dissipation (g), calculated at $t\simeq1.2\mu\mathrm{s}$(d-e) and $t\simeq2.5\mu\mathrm{s}$(f-g).
  }
  \label{fig:CatStates}
\end{figure}

We benchmark our protocol for different values of the Gilbert damping parameter $\alpha_{G}$, as shown in Fig.~\ref{fig:CatStates}, using drive parameters $\phi_{\mathrm{ac}}=\pi/10$, $\omega_{\mathrm{ac}}=\omega_{m}$ and circuit parameters from Fig.~\ref{fig:couplings}.
We first address the entanglement of the bipartite system by monitoring the evolution of the logarithmic negativity $E_N=\log_2(2N(\rho)+1)$, where $N(\rho)$ is the sum of negative eigenvalues of the partial transpose of the joint density matrix $\rho$~\cite{vidal2002computable}.
Fig.~\ref{fig:CatStates}(a) shows that after qubit initialization and evolution under $\hat{\widetilde{H}}$, the magnon-qubit entanglement increases up to a maximum value ($E_N^\mathrm{max}=1$ without dissipation) before the magnon decay takes over and eventually destroys it.

As explained above, collapsing the qubit in ${\vert 0\rangle_q}$ prepares a magnonic \emph{even cat state}.
In Fig.~\ref{fig:CatStates}(b) we plot the cat ``size'' ${S=\left\vert\beta(t)\right\vert^2 }$, defined as the squared distance between the superposed coherent states in phase-space~\cite{deleglise2008reconstruction,vlastakis2013deterministically}.
Fig.~\ref{fig:CatStates}(c) shows the prepared state fidelity $\mathcal{F}=\sqrt{\langle\psi_\mathrm{even}\vert\rho_m\vert\psi_\mathrm{even}\rangle}$~\cite{nielsen2010quantum,johansson2012qutip}, where $\rho_m$ is the magnon density matrix after tracing out the qubit.
We can visualize the cat state in terms of the Wigner quasi-probability distribution, ${W(\alpha)=2/\pi\Tr{D^\dagger(\alpha)\rho_m D(\alpha)e^{i\pi \hat{m}^\dagger\hat{m}}}}$, where ${D(\alpha)=e^{\alpha\hat{m}^\dagger-\alpha^{*}\hat{m}}}$ is the magnon displacement operator~\cite{cahill1969density}.
In Figures~\ref{fig:CatStates}(d)-\ref{fig:CatStates}(f) we plot \(W(\alpha)\) for $\alpha_{G}=10^{-4}$ and $\alpha_{G}=10^{-5}$, calculated at selected times where \(\mathcal{F}\simeq80~\%\) (d,f) and \(\mathcal{F}\simeq95~\%\) (e).
Fig.~\ref{fig:CatStates}(g) shows the Wigner function at $t=2.5~\mathrm{\mu s}$ for the dissipationless case.
We clearly observe the characteristic quantum features of cat states, such as interference fringes with \(W(\alpha)<0\).
It should therefore be possible to prepare high-fidelity large cat states with sizes \(S \ge 40\) for realistic values of ${10^{-5}\le\alpha_{G}\le10^{-4}}$~\cite{tabuchi2015coherent,klingler2017gilbert,rameshti2021cavity}.
Our analysis holds for non-interacting magnons.
When the total magnon number $\langle \hat{m}^\dagger \hat{m}\rangle=\left\vert\beta(t)\right\vert^2/2$ is much smaller than the total number of spins ($N_S$) magnon-magnon interactions are negligibly small~\cite{elyasi2020resources}.
In our set-up corrections would be necessary when $\left\vert \tilde{g}_{\mathrm{rp}} t \right\vert^2/2 \sim N_S$.
With parameters $\tilde{g}_{\mathrm{rp}}/(2\pi)\sim 0.5~\mathrm{MHz}$ and $N_S\sim10^{12}$ pumping such  magnon numbers takes seconds, which is orders of magnitude larger than their lifetime.

The quantum nature of the magnetic states can be verified by homodyne magnetization state tomography of $W(\alpha)$~\cite{hioki2021state} by combining ac spin pumping with inverse spin Hall voltage measurements.
Wigner tomography can also achieved by operating the magnon-qubit system in the strong dispersive regime, $J^2/(\omega_q-\omega_m)\gg\kappa,1/T_1$, and using the transmon as a magnon detector~\cite{vlastakis2013deterministically,vlastakis2015characterizing,langford2017experimentally}.
Finally, rapid modulation of the radiation-pressure coupling combined with magnon displacement operations and qubit measurements can yield signatures of the prepared cat state~\cite{asadian2014probing}.
The last two methods require a dispersively coupled microwave resonator to the qubit~\cite{Schuster2007}.

\emph{Conclusion.}~Strong and tunable magnon-qubit couplings can be realized in a hybrid quantum system comprising a magnetized YIG sphere directly coupled via magnetic flux to a transmon qubit in a planar superconducting circuit.
This architecture features both resonant magnon-qubit exchange as well as purely nonlinear interactions, in a flexible and compact geometry with in-situ tunability.
To the best of our knowledge, we are the first to propose a device that employs the direct interaction of magnons with superconducting qubits and, in particular, nonlinear magnon-qubit couplings of the radiation-pressure type.
The intrinsic nonlinearity of both the transmon qubit and the radiation-pressure coupling empowers the creation of nontrivial quantum magnonic states even in the weak excitation regime, in which magnons behave as harmonic oscillators.
We devised and tested an analog protocol that is particularly useful for the high-fidelity generation of large macroscopic quantum superpositions of magnetization under realistic experimental conditions.
Our results enrich the quantum control toolbox in magnonic devices and open new possibilities for constructing artificial 2D quantum magnonic networks~\cite{nisoli2013artificial,rusconi2019hybrid} or building magnonic analogs of the bosonic codes paradigm~\cite{mirrahimi2014dynamically,chamberland2020building,weigand2019realizing} with transmons playing the role of ancillary control elements.

\emph{Acknowledgments.}~We thank Christian Dickel and Mehrdad Elyasi for reading and commenting on the manuscript.
This research was supported by the Dutch Foundation for Scientific Research (NWO) and the JSPS Kakenhi grant no. 19H00645.


%
\cleardoublepage
\pagebreak
\newpage

\renewcommand{\theequation}{S\arabic{equation}}
\renewcommand{\thefigure}{S\arabic{figure}}
\renewcommand{\thetable}{S\arabic{table}}
\renewcommand{\bibnumfmt}[1]{[S#1]}
\onecolumngrid

\begin{center}
{\Large {\bf Supplementary Information}}
\end{center}

\vspace{10pt}

\section{Derivation of the coupled magnon-qubit system Hamiltonian}

We consider a transmon qubit, formed by a SQUID loop in parallel to a capacitor $C$, in the presence of an external flux $\Phi_{\mathrm{ext}}$.
Its Hamiltonian reads
\begin{equation}
\hat{H}_{\mathrm{circ}}=4E_{C}\hat{N}^{2}-E_{J}^{\mathrm{max}}S(\phi_{\mathrm{ext}})\cos{\left(  \hat{\delta}-\arctan{(}{a_{J}}{\tan{\phi
}_{\mathrm{ext}})}\right)  } ,
\label{eq:CircuitHamiltonian}
\end{equation}
where $\phi_{\mathrm{ext}}=\pi\Phi_{\mathrm{ext}}/\Phi_{0}$, ${S(\phi_{b})=\sqrt{\cos^{2}{\phi_{b}}+a_{J}^{2}\sin^{2}{\phi_{b}}}}$, and $E_{C}=e^2/(2C),~E_{J}^{\mathrm{max}}=E_{J}^{1}+E_{J}^{2}$ is the charging and the maximum Josephson energy of the transmon, respectively, while $a_{J}=\vert E_{J}^{1}-E_{J}^{2}\vert /E_{J}^{\mathrm{max}}$ is the SQUID asymmetry ($E_{J}^{1},~E_{J}^{2}$ are the individual Josephson energies of each junction).
The canonically conjugate operators $\hat{\delta}$,~$\hat{N}$ describe the superconducting phase difference and the number of tunneling Cooper pairs across the SQUID , respectively.
The above Hamiltonian can be simplified by the trigonometric relations
\begin{align}
\cos{\left(  x-y\right)  } &  =\cos{x}\cos{y}+\sin{x}\sin{y} ,\nonumber\\
\cos\left(  \arctan{x}\right)   &  =1/\sqrt{1+x^{2}} ,\\
\sin\left(  \arctan{x}\right)   &  =x/\sqrt{1+x^{2}} ,\nonumber
\end{align}
 to
\begin{equation}
\hat{H}_{\mathrm{circ}}=4E_{C}\hat{N}^{2}-E_{J}^{\mathrm{max}}|\cos{\phi
}_{\mathrm{ext}}|\left(  \cos{\hat{\delta}}+a_{J}\tan{\phi}_{\mathrm{ext}}\sin{\hat{\delta}}\right)  .
\end{equation}

The YIG sphere is placed at an in-plane distance $d$ from the closest point in the SQUID loop.
The \textquotedblleft Kittel\textquotedblright\ mode is a precession of the macrospin $\bm{{S}}$ with frequency $\omega_{m}=\gamma_{0}B_{z}$~\cite{tabuchi2016quantum,lachance2019hybrid,rameshti2021cavity} around a magnetic field $B_{z}$ along the in-plane z-direction, where $\gamma_0$ is the gyromagnetic ratio.
The Holstein-Primakoff transformation $\hat{S}_{x}=\sqrt{\frac{S}{2}}(\hat{m}+\hat{m}^{\dagger})$, $\hat{S}_{y}=i\sqrt{\frac{S}{2}}(\hat{m}-\hat{m}^{\dagger})$, $\hat{S}_{z}=S-\hat{m}^{\dagger}\hat{m}$ of a spin $S$ leads to the Hamiltonian
\begin{equation}
\hat{H}_{0}^{M}=\hbar\omega_{m}\hat{m}^{\dagger}\hat{m} ,
\end{equation}
where $\hat{m}^{(\dagger)}$ are bosonic operators describing the annihilation (creation) of a single magnon, in the limit of weak excitation or $\langle\hat{m}^{\dagger}\hat{m}\rangle\ll S$~\cite{tabuchi2016quantum,lachance2019hybrid,rameshti2021cavity}.

The magnetic moment $\bm{\hat{\mu}}$ generates a magnetic stray field $\mathbf{B}_{\text{YIG}}(\bm{\hat{\mu}})$, which induces a magnetic flux through the SQUID loop area A,
\begin{equation}
\Phi(\bm{\hat{\mu}})=\iint{\mathbf{B}_{\text{YIG}}(\bm{\hat{\mu}}) \cdot d\mathbf{{A}}} .
\label{eq:FluxIntegral}
\end{equation}
The induced flux can be expressed as $\Phi(\bm{\hat{\mu}})=\Phi(\langle\hat{\mu}\rangle_0)+\Phi(\Delta\bm{\hat{\mu}})$, where $\Phi(\Delta\bm{\hat{\mu}})$ is the flux induced by the quantum fluctuations of the magnetic moment $\Delta\bm{\hat{\mu}}$ and $\Phi(\langle\hat{\mu}\rangle_0)$ is a constant flux offset do to the equilibrium value of the magnetic moment $\langle\hat{\mu}\rangle_0$.
The latter can be compensated by an externally applied DC flux bias $\Phi_{b}$, therefore we can define the total external flux
\begin{equation}
\Phi_{\mathrm{ext}}=\Phi_{b}+\Phi(\Delta\bm{\hat{\mu}}) .
\end{equation}
The Cartesian components of $\Delta\bm{\hat{\mu}}$ contribute to the flux as
\begin{equation}
\phi(\Delta\bm{\hat{\mu}})=\pi\Phi(\Delta\bm{\hat{\mu}})/\Phi_{0}=\frac{\mu_{0}}{4\Phi_{0}d_{\mathrm{min}}}(I_{x}\Delta\hat{\mu}_{x}+I_{y}\Delta\hat{\mu}_{y}+I_{z}\Delta\hat{\mu}_{z}) ,
\end{equation}
where $d_{\mathrm{min}}$ is the minimum distance between center of the sphere and the closest point of the SQUID loop, and
\begin{equation}
\Delta\hat{\mu}_{x}=\mu_{\mathrm{zpf}}(\hat{m}+\hat{m}^{\dagger}) ,~\Delta\hat{\mu}_{y}=i\mu_{\mathrm{zpf}}(\hat{m}-\hat{m}^{\dagger}) ,~\Delta\hat{\mu}_{z}=\hbar\gamma_{0}\hat{m}^{\dagger}\hat{m} .
\label{eq:MagnonOperatorsSI}
\end{equation}
with transverse magnetic zero-point fluctuation amplitude $\mu_{\mathrm{zpf}}=\hbar\gamma_{0}\sqrt{N_S/2}$, where $N_S$ is the total number of spins.
The dimensionless factors $I_{x},I_{y},I_{z}$ arise from the integration in Eq.~(\ref{eq:FluxIntegral}).
In the case where the magnet is placed inside the loop, the spatial distribution of the superconducting order parameter should also be considered in the calculations, e.g., as in Ref.~\cite{rybakov2022flux}.

The SQUID in Fig.~\ref{fig:scheme} of the main text is a large ring ($R_{\mathrm{SQUID}}\gg d$), so $I_{y}=0$ and ${I_{x,z}\simeq-1}$ for $d=R_{\mathrm{YIG}}$. ${\vert I_{x}\vert}$ is largest when its center is in the plane of the SQUID~\cite{rusconi2019hybrid}, so we expect that an in-plane magnetic disk with the same radius gives very similar results.
We disregard a very small contribution from $\Delta\hat{\mu}_{z}$. 
For ${R_{\mathrm{YIG}}=d=3~\mu\mathrm{m}}$ and typically ${R_{\mathrm{SQUID}}\gtrsim10\mathrm{\mu m}}$~\cite{langford2017experimentally,kounalakis2019nonlinear}, the flux induced by the quantum spin fluctuations
\begin{equation}
\phi(\Delta\bm{\hat{\mu}})=\frac{I_{x}\mu_{0}}{4\Phi_{0}d_{\mathrm{min}}}\Delta\hat{\mu}_{x}  ,
\label{eq:Phi_x}
\end{equation}
so  $\phi(\Delta\bm{\hat{\mu}})\ll1$ for $d_{\mathrm{min}}=\sqrt{2}d$.

With $\phi(\bm{\hat{\mu}})\ll\phi_{b}$, Eq.~(\ref{eq:CircuitHamiltonian}) becomes
\begin{equation}
\hat{H}_{\mathrm{circ}}=4E_{C}\hat{N}^{2}-sE_{J}^{\mathrm{max}}\left([\cos{\phi_{b}}-\phi(\Delta\bm{\hat{\mu}})\sin{\phi_{b}}]\cos{\hat{\delta}}+a_{J}[\sin{\phi_{b}}+\phi(\Delta\bm{\hat{\mu}})\cos{\phi_{b}}]\sin{\hat{\delta}}\right)  ,
\end{equation}
where $s\dot{=}\mathrm{sgn}[\cos{\phi_{\text{ext}}}]$ can be simplified to $s=\text{sgn}[\cos${$\phi_{b}$}$]$ except for a small interval around $\Phi_{\text{b}}=(2k+1)\Phi_{0}/2$ of the order of $\Phi(\Delta\bm{\hat{\mu}})=10^{-3}\Phi_{0}$ for the parameters considered here.

The total system Hamiltonian is
\begin{align}
\hat{H} &  =\hat{H}_{0}^{M}+\hat{H}_{0}^{T}+\hat{H}_{\mathrm{int}} ,\\
\hat{H}_{0}^{T} &  =4E_{C}\hat{N}^{2}-E_{J}^{\mathrm{max}}\sqrt{\cos^{2}{\phi_{b}}+a_{J}^{2}\sin^{2}{\phi_{b}}}\cos{\left(  \hat{\delta}-\arctan{(a_{J}\tan{\phi_{b}})}\right)  } ,\\
\hat{H}_{\mathrm{int}} &  =sE_{J}^{\mathrm{max}}\phi(\Delta\bm{\hat{\mu}})\left[  \sin{\phi_{b}}\cos{\hat{\delta}}-a_{J}\cos{\phi_{b}}\sin{\hat{\delta}}\right]  ,
\end{align}
where $\hat{H}_{0}^{\text{T}}$ ($\hat{H}_{0}^{\text{M}}$) is the bare transmon (magnon) Hamiltonian.
In terms of the transmon phase operator ${\hat{\tilde{\delta}}=\hat{\delta}-\arctan{(a_{J}\tan{\phi_{b}})}}$ the interaction Hamiltonian reads
\begin{equation}
\hat{H}_{\mathrm{int}}=sE_{J}^{\mathrm{max}}\phi(\Delta\bm{\hat{\mu}})\left[  \sin{\phi_{b}}\cos{\left(  \hat{\tilde{\delta}}+\arctan{(a_{J}\tan{\phi_{b}})}\right)  }-a_{J}\cos{\phi_{b}}\sin{\left(  \hat{\tilde{\delta}}+\arctan{(}{a_{J}}{\tan{\phi_{b}})}\right)}\right] ,
\end{equation}
and can be further simplified to
\begin{equation}
\hat{H}_{\mathrm{int}}=\frac{E_{J}^{\mathrm{max}}\phi(\Delta\bm{\hat{\mu}})}{S(\phi_{b})}\left[  \frac{\sin{2\phi_{b}}}{2}\left(  1-a_{J}\right)\cos{\hat{\tilde{\delta}}}-a_{J}\sin{\hat{\tilde{\delta}}}\right]  .
\label{eq:Hint_total}
\end{equation}

In terms of the bosonic field operators $\hat{c}^{(\dagger)}$ of the qubit excitation 
\begin{equation}
\hat{N}=i\left(  \frac{E_{J}^{\mathrm{max}}S(\phi_{b})}{32E_{C}}\right)^{1/4}(\hat{c}^{\dagger}-\hat{c}),~\hat{\tilde{\delta}}=\left(  \frac{2E_{C}}{E_{J}^{\mathrm{max}}S(\phi_{b})}\right)  ^{1/4}(\hat{c}+\hat{c}^{\dagger}). \label{eq:TransmonOperatorsSI}
\end{equation}
When $E_{J}^{\mathrm{max}}S(\phi_{b})\gg E_{C}$, i.e., the transmon regime, the zero-point fluctuations of the phase variable are small, ${\tilde{\delta}}_{\mathrm{zpf}}=\left(  \frac{2E_{C}}{E_{J}^{\mathrm{max}}S(\phi_{b})}\right)  ^{1/4}\ll1$, and the transmon is well-described by the Duffing oscillator Hamiltonian,
\begin{align}
\hat{H}_{0}^{T} &  =4E_{C}\hat{N}^{2}+E_{J}^{\mathrm{max}}S(\phi_{b})\left(\frac{\hat{\tilde{\delta}}^{2}}{2}-\frac{\hat{\tilde{\delta}}^{4}}{24}\right)\nonumber\\
&  =\hbar\omega_{q}\hat{c}^{\dagger}\hat{c}-\frac{E_{C}}{2}\hat{c}^{\dagger}\hat{c}^{\dagger}\hat{c}\hat{c},
\end{align}
where $\omega_{\text{q}}~=\left(  \sqrt{8E_{C}E_{J}^{\mathrm{max}}S(\phi_{b})}-E_{C}\right)  /\hbar$ is the transmon qubit frequency.

\begin{figure}[h]
  \begin{center}
  \includegraphics[width=0.7\linewidth]{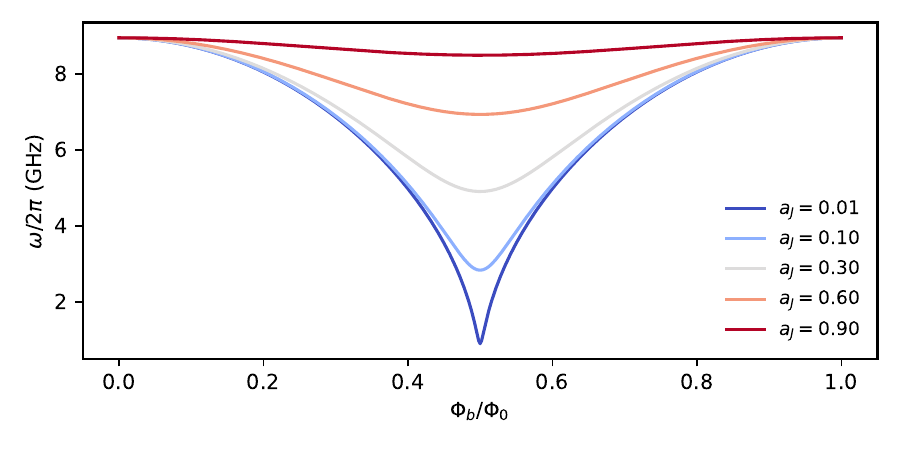}
  \end{center}
  \caption{
  { Transmon excitation frequency vs. $\Phi_b$ and $\alpha_J$ for the same transmon parameters as used in Fig.~\ref{fig:couplings} of the main text, i.e., $E_{J}^{\text{max}}/h=50$~GHz and $E_{C}/h=200$~MHz}. 
  }
  \label{fig:FreqVflux}
\end{figure}

Using the expression from Eq.~(\ref{eq:Phi_x}) for ${I_x=-1}$ and expanding up to $\mathcal{O}[\hat{\tilde{\delta}}^{4}]$ in the interaction Hamiltonian we find
\begin{equation}
\hat{H}_{\mathrm{int}}=\frac{\mu_{0}E_{J}^{\mathrm{max}}}{4\Phi_{0}d_{\mathrm{min}}S(\phi_{b})}\left[  \frac{\sin{2\phi_{b}}}{2}\left(1-a_{J}^{2}\right)  \Delta\hat{\mu}_{x}\left( -1+ \frac{\hat{\tilde{\delta}}^{2}}{2}-\frac{\hat{\tilde{\delta}}^{4}}{24}\right)  +a_{J}\left(  \hat{\tilde{\delta}}-\frac{\hat{\tilde{\delta}}^{3}}{6}\right)  \Delta\hat{\mu}_{x}\right]  .
\end{equation}
In terms of the annihilation (creation) operators $\hat{c}^{(\dagger)},\hat{m}^{(\dagger)}$ the above expression reads
\begin{equation}
\hat{H}_{\mathrm{int}}=\hat{H}_{\mathrm{rp}}+\hat{H}_{J}+\hat{H}^{\left(\Delta\hat{\mu}_{x}\hat{\tilde{\delta}}^{3}\right)  }+\hat{H}^{\left(\Delta\hat{\mu}_{x}\hat{\tilde{\delta}}^{4}\right)  }+\hat{H}_D^M
,\label{eq:Hint_total_expanded}
\end{equation}
where
\begin{align}
\hat{H}_{\mathrm{rp}}/\hbar &  =g_{\mathrm{rp}}\hat{c}^{\dagger}\hat{c}(\hat{m}+\hat{m}^{\dagger}),\\
\hat{H}_{J}/\hbar &  =~J~(\hat{c}^{\dagger}\hat{m}+\hat{c}\hat{m}^{\dagger}),
\end{align}
are the leading terms in the order of ($\mathcal{O}[\Delta\hat{\mu}_{x}\hat{\tilde{\delta}}^{2}]$ and $\mathcal{O}[\Delta\hat{\mu}_{x}\hat{\tilde{\delta}}]$) after expressing $\Delta\hat{\mu}_{x}$ and $\hat{\tilde{\delta}}$ in terms of the field operators~(\ref{eq:MagnonOperatorsSI}) and~(\ref{eq:TransmonOperatorsSI}), and applying the rotating-wave approximation (RWA) by disregarding fast rotating terms, like $(\hat{c}^{\dagger})^{n}\hat{m}^{(\dagger)}$.
$\hat{H}_{\mathrm{rp}}$ is a radiation-pressure type interaction with
\begin{equation}
g_{\mathrm{rp}}=\frac{\mu_{0}\mu_{\mathrm{zpf}}}{16d_{\mathrm{min}}\Phi_{0}}\frac{\sqrt{8E_{J}^{\mathrm{max}}E_{C}}(1-a_{\text{J}}^{2})\sin{2\phi_{b}}}{[S(\phi_{b})]^{3/2}},
\label{eq:gRPcouplingSI}
\end{equation}
while $\hat{H}_{J}$ describes a qubit-magnon exchange scattering with
\begin{equation}
J=\frac{\mu_{0}\mu_{\mathrm{zpf}}}{4d_{\mathrm{min}}\Phi_{0}}\frac{a_{J}\left(  2E_{C}(E_{J}^{\mathrm{max}})^{3}\right)  ^{1/4}}{[S(\phi_{b})]^{5/4}}.
\end{equation}

The term
\begin{equation}
\hat{H}_D^M=-\frac{\mu_{0}E_{J}^{\mathrm{max}}}{8\Phi_{0}d_{\mathrm{min}}S(\phi_{b})}\sin{2\phi_{b}}\left(1-a_{J}^{2}\right)  \Delta\hat{\mu}_{x}=-\frac{g_{\mathrm{rp}}}{\delta_{\mathrm{zpf}}^2}(\hat{m}+\hat{m}^{\dagger})
\end{equation}
is a constant magnon displacement that depends on the zero-point fluctuations of the superconducting phase difference.
This term does not affect the interaction dynamics and can be canceled by a coherent displacement operation $D(\alpha)=e^{\alpha\hat{m}^\dagger-\alpha^{*}\hat{m}}$, with $\alpha=\frac{ig_{\mathrm{rp}}}{\delta_{\mathrm{zpf}}^2}t$, on the magnon state.
By the same logic, we drop a similar term in the radiation-pressure interaction that originates from $\hat{\tilde{\delta}}^{2}\Delta\hat{\mu}_{x} \sim (\hat{c}+\hat{c}^{\dagger})^2(\hat{m}+\hat{m}^{\dagger})$.

The interaction terms of the order of $\Delta\hat{\mu}_{x}\hat{\tilde{\delta}}^{3}$ and $\Delta\hat{\mu}_{x}\hat{\tilde{\delta}}^{4}$ in Eq.~(\ref{eq:Hint_total_expanded}) 
are small compared to $\hat{H}_{\mathrm{rp}},~\hat{H}_{J}$ in the transmon regime; we analyze them here for the sake of completeness.
In the RWA,
\begin{align}
\hat{H}^{\left(  \Delta\hat{\mu}_{x}\hat{\tilde{\delta}}^{3}\right)  }&=-\frac{\mu_{0}E_{J}^{\mathrm{max}}}{4\Phi_{0}d_{\mathrm{min}}S(\phi_{b})}a_{J}\Delta\hat{\mu}_{x}\frac{\hat{\tilde{\delta}}^{3}}{6}\nonumber\\
&=~J^{\prime}\hbar~\left[  (\hat{c}^{\dagger}\hat{m}+\hat{c}\hat{m}^{\dagger})~+~\left(\hat{c}^{\dagger}\hat{m}(\hat{c}^{\dagger}\hat{c})+(\hat{c}^{\dagger}\hat{c})\hat{m}^{\dagger}\hat{c}\right)  \right]  ,
\label{eq:exchangecorrection}
\end{align}
yields a qubit-magnon exchange interaction and an additional \emph{correlated exchange} term with coupling strength
\begin{equation}
J^{\prime}=-\frac{J}{2}\sqrt{\frac{2E_{C}}{E_{J}^{\mathrm{max}}S(\phi_{b})}},
\end{equation}
which is small in the transmon regime $E_{J}^{\mathrm{max}}S(\phi_{b})\gg E_{C}$.
The first term in Eq.~(\ref{eq:exchangecorrection}) is a correction to the exchange coupling $J\rightarrow J+J^{\prime}$, while the second term describes a correlated exchange interaction that becomes important when the higher transmon states are excited.

Following the same procedure,
\begin{align}
\hat{H}^{\left(  \Delta\hat{\mu}_{x}\hat{\tilde{\delta}}^{4}\right)  }&=-\frac{\mu_{0}E_{J}^{\mathrm{max}}}{4\Phi_{0}d_{\mathrm{min}}S(\phi_{b})}\frac{\sin{2\phi_{b}}}{2}\left(  1-a_{J}^{2}\right)  \Delta\hat{\mu}_{x}\frac{\hat{\tilde{\delta}}^{4}}{24}\nonumber\\
&=\hbar g_{\mathrm{rp}}^{\prime}\left(  \hat{c}^{\dagger}\hat{c}+\frac{1}{2}\hat{c}^{\dagger}\hat{c}^{\dagger}\hat{c}\hat{c}\right)  (\hat{m}+\hat{m}^{\dagger}),
\end{align}
where
\begin{equation}
g_{\mathrm{rp}}^{\prime}=-\frac{g_{\mathrm{rp}}}{2}\sqrt{\frac{2E_{C}}{E_{J}^{\mathrm{max}}S(\phi_{b})}}.
\end{equation}
This term corrects the radiation-pressure $g_{\mathrm{rp}}\rightarrow g_{\mathrm{rp}}+g_{\mathrm{rp}}^{\prime}$ which is not relevant in the single-excitation manifold of the qubit.

\begin{figure}[h]
  \begin{center}
  \includegraphics[width=0.7\linewidth]{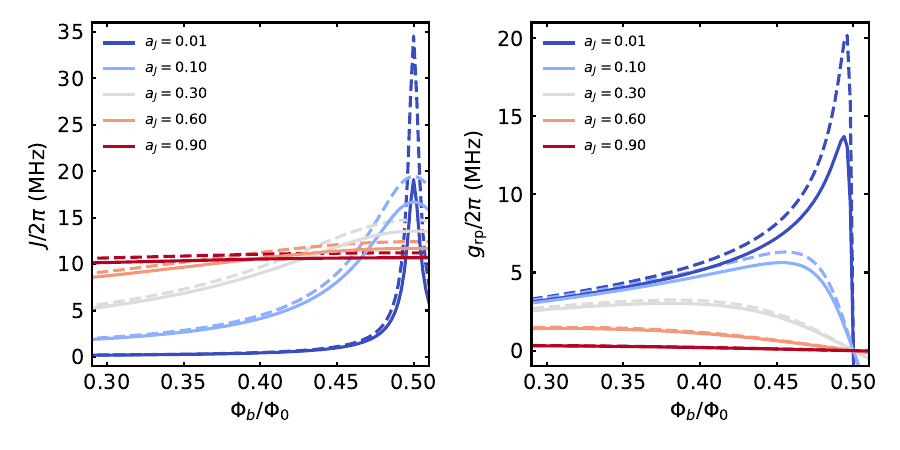}
  \end{center}
  \caption{
  { Couplings $J$ and $g_{\mathrm{rp}}$ vs $\Phi_b$ calculated up to leading order $\mathcal{O}[\Delta\hat{\mu}_{x}\hat{\tilde{\delta}}^{2}]$ (dashed curves) and their corrections $J+J^\prime$ and $g_{\mathrm{rp}}+g_{\mathrm{rp}}^\prime$ including the higher-order terms (solid) as in Fig.~\ref{fig:couplings}}. 
  }
  \label{fig:CouplingsCorr}
\end{figure}

We calculated both correction terms in the simulations described in the main text, but did not find noticeable effects on the system dynamics. Fig.~\ref{fig:CouplingsCorr}
shows that they become significant when operating beyond the transmon regime, i.e., for $a_J\rightarrow0$ and $\Phi_b\rightarrow\Phi_0/2$, however.

\section{The role of junction capacitance in circuit quantization}

In the Hamiltonian of the flux-tunable transmon in Eq.~(\ref{eq:CircuitHamiltonian}) we assumed that the superconducting phase across the SQUID and the external flux are given by $\delta~=~(\delta_{1}-\delta_{2})/2$ and ${2\pi\Phi_{\mathrm{ext}}/\Phi_0~=~(\delta_{1}+\delta_{2})}$, respectively, where $\delta_{1,2}$ denote the phase differences across each Josephson junction. However, in the presence of time-dependent fluxes $\Phi_{\mathrm{ac}}(t)$ this is not true in general ~\cite{you2019circuit, riwar2021circuit} .
A unique linear combination of $\delta_{1,2}$ defines $\delta$ (in the so-called \emph{irrotational gauge}) as a function of the effective junction capacitances, $C_{1,2}$, as
\begin{equation}
\delta_{1}=\delta+\frac{C_{2}}{C_{\Sigma}}2\pi\Phi_{\mathrm{ac}}(t)/\Phi_0,~\delta_{2}=-\delta+\frac{C_{1}}{C_{\Sigma}}2\pi\Phi_{\mathrm{ac}}(t)/\Phi_0 ,
\end{equation}
with $C_{\Sigma}=C_{1}+C_{2}$.
The inductive energy of the SQUID is then~\cite{zorin1996quantum}
\begin{align}
H_{\mathrm{ind}} &  =~-E_{J}^{\mathrm{max}}\left[  \left(  \frac{1+a_{J}}{2}\right)  \cos{\delta_{1}}+\left(  \frac{1-a_{J}}{2}\right)  \cos{\delta_{2}}\right]  \nonumber\\
&  =~-E_{J}^{\mathrm{max}}\sqrt{\cos^{2}\frac{\delta_{1}+\delta_{2}}{2}+a_{J}^{2}\sin^{2}\frac{\delta_{1}+\delta_{2}}{2}}\cos{\left[  \frac{\delta_{1}-\delta_{2}}{2}-\arctan{\left(  a_{J}\tan\frac{\delta_{1}+\delta_{2}}{2}\right)  }\right]  }.
\end{align}
Adding the constant $\phi_{b}$ and imposing the fluxoid quantization condition${(\delta_{1}+\delta_{2})/2=\phi_{\text{ext}}}=(\phi_{b}+\phi_{\mathrm{ac}}(t))$, where here $\phi_{\mathrm{ac}}(t)=\phi(\Delta\bm{\hat{\mu}})$,
\begin{equation}
H_{\mathrm{ind}}=-E_{J}^{\mathrm{max}}\sqrt{\cos^{2}\phi_{\mathrm{ext}}+a_{J}^{2}\sin^{2}\phi_{\mathrm{ext}}}\cos{\left[  \left(  \hat{\delta}+\frac{C_{\Delta}}{C_{\Sigma}}\phi(\Delta\bm{\hat{\mu}})\right)-\arctan{\left(  a_{J}\tan\phi_{\mathrm{ext}}\right)  }\right]  },
\end{equation}
with $C_{\Delta}=C_{2}-C_{1}$.
This expression agrees with Eq.~(\ref{eq:CircuitHamiltonian}) after replacing $\hat{\delta}\rightarrow\hat{\delta}+\frac{C_{\Delta}}{C_{\Sigma}}\phi(\Delta\bm{\hat{\mu}})$.

\begin{figure}[h]
  \begin{center}
  \includegraphics[width=0.4\linewidth]{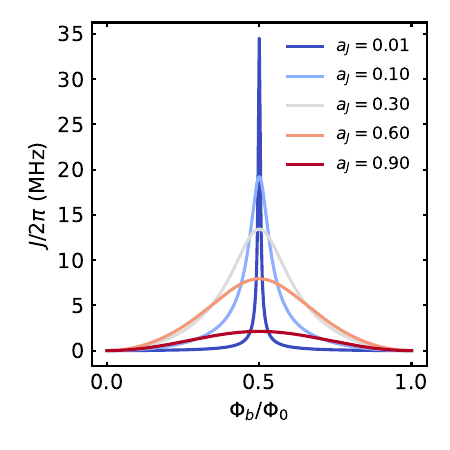}
  \end{center}
  \caption{
  { Qubit-magnon exchange coupling from Eq.~(\ref{eq:Jcoupling_SI}) when the asymmetry of the capacitances $C_\Delta/C_\Sigma$ matches the SQUID asymmetry $a_J$.} 
  }
  \label{fig:Jcoupling_SI}
\end{figure}

The modified interaction Hamiltonian in Eq.~(\ref{eq:Hint_total}) reads
\begin{equation}
H_{\mathrm{int}}=\frac{E_{J}^{\mathrm{max}}\phi(\Delta\bm{\hat{\mu}})}{S(\phi_{b})}\left[  \frac{\sin{2\phi_{b}}}{2}\left(  1-a_{\text{J}}^{2}\right)  \cos{\hat{\tilde{\delta}}}-\left(  a_{J}-\frac{C_{\Delta}}{C_{\Sigma}}[S(\phi_{b})]^{2}\right)  \sin{\hat{\tilde{\delta}}}\right]  .
\end{equation}
While the radiation-pressure remains unaffected, the modified qubit-magnon exchange coupling is
\begin{equation}
J=\frac{\mu_{0}\mu_{\mathrm{zpf}}}{4d_{\mathrm{min}}\Phi_{0}}\left(a_{J}-\frac{C_{\Delta}}{C_{\Sigma}}[S(\phi_{b})]^{2}\right)  \frac{\left(2E_{C}(E_{J}^{\mathrm{max}})^{3}\right)  ^{1/4}}{[S(\phi_{b})]^{5/4}},
\label{eq:Jcoupling_SI}
\end{equation}
which reduces to Eq.~(\ref{eq:Jcoupling}) for $C_{1}=C_{2}$.
Note that $C_{1,2}$ are effective parameters that may differ from the nominal junction capacitances (and even change sign); their actual values depend on the specific magnetic stray fields distribution and the distance between both junctions and the magnetic dipole~\cite{riwar2021circuit}.
Therefore, $C_{1,2}$ and the modified exchange coupling can be estimated for specific experimental designs~\cite{riwar2021circuit}.

In Fig.~\ref{fig:Jcoupling_SI} we plot the qubit-magnon exchange coupling from Eq.~(\ref{eq:Jcoupling_SI}) for $C_\Delta/C_\Sigma=a_J$. The coupling vanishes at $\Phi_b=k\Phi_0~(k\epsilon\mathbb{Z})$ for all values of the asymmetry and when the loop becomes a single junction, e.g. $E_J^1=C_1=0$.

\section{Critical distance}
Our coupling diverges when the sphere touches the SQUID loop, which signals a breakdown of the point dipole approximation. However, at close proximity the stray fields exceed the critical field that destroys the superconducting order.
Here, we estimate the minimum allowed distance $d_{\text{c}}$ between the magnet and the SQUID.
to be that at which the stray field equals the critical magnetic field $B_{c}$ of the superconducting wire with thickness \(d_w\)~\cite{rusconi2019hybrid}
\begin{equation}
d_{c}=\frac{d_w}{2}+\left(  \frac{2\mu_{0}M_{s}}{3B_{c}}\right)^{1/3}R_{\mathrm{YIG}},
\end{equation}
where $M_{s}$ is the saturation magnetization with $M_s^{\mathrm{YIG}}\sim2\times10^{5}$~A/m$^{3}$~\cite{klingler2017gilbert}.

\begin{figure}[h]
  \begin{center}
  \includegraphics[width=0.5\linewidth]{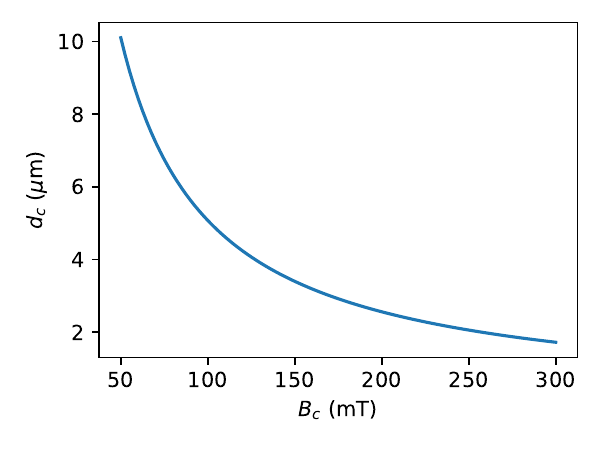}
  \end{center}
  \caption{
  {Critical distance of a YIG sphere that destroys the superconducting order in the SQUID.} 
  }
  \label{fig:CriticalField}
\end{figure}

In Fig.~\ref{fig:CriticalField} we plot $d_{c}$ vs. $B_{c}$ for a sphere radius $R_{\mathrm{YIG}}=3\,\mathrm{\mu m}$, assuming $d_w=100$~nm.
A distance from the center of the magnet to the SQUID loop of $d_{\mathrm{min}}=\sqrt{2}R_{\mathrm{YIG}}\simeq4.2~\mu$m would allow to use superconductors with critical field down to $120\,$mT,
which is two orders of magnitude smaller than the critical magnetic field of NbTiN~\cite{popinciuc2012zero} used in highly coherent transmon qubit implementations~\cite{langford2017experimentally,kounalakis2019nonlinear,murray2021material}.
Josephson junctions are typically made from aluminum and for a typical junction thickness $\sim10$~nm~\cite{kounalakis2019nonlinear} the critical field is around 1$\,$T~\cite{krause2021magnetic}.
In Fig.~\ref{fig:scheme}, the distance of the magnet from each junction is $\sqrt{R_{\mathrm{YIG}}^{2}+d^{2}+R_{\mathrm{SQUID}}^{2}}~\simeq~11 \, \mu$m, with a stray field of $50\,$mT, which does not affect
the coherence of the qubit~\cite{krause2021magnetic}.

\section{Temperature}
\begin{figure}[h]
  \begin{center}
  \includegraphics[width=0.6\linewidth]{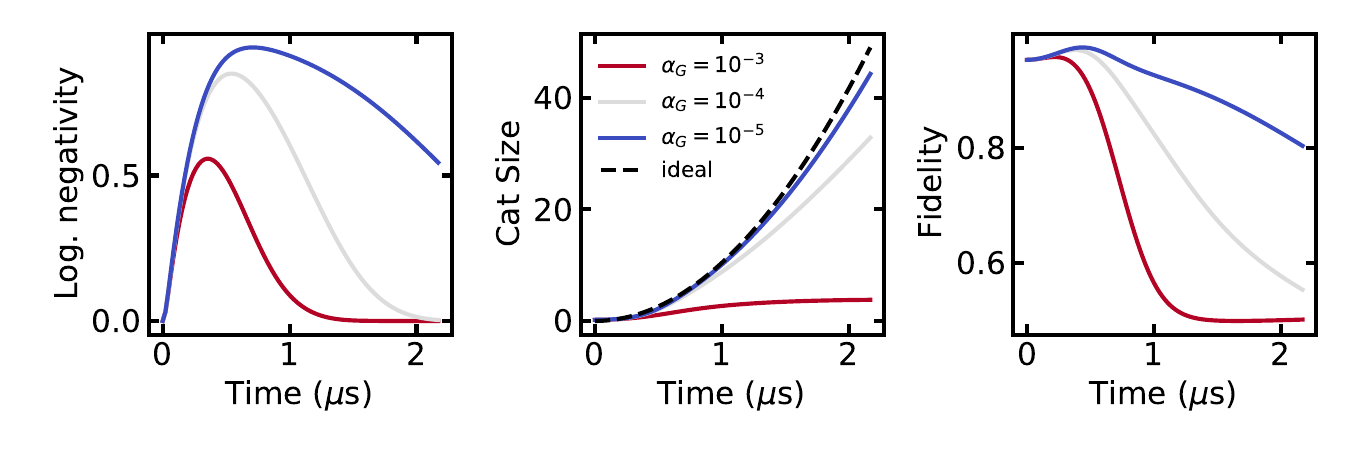}
  \includegraphics[width=0.6\linewidth]{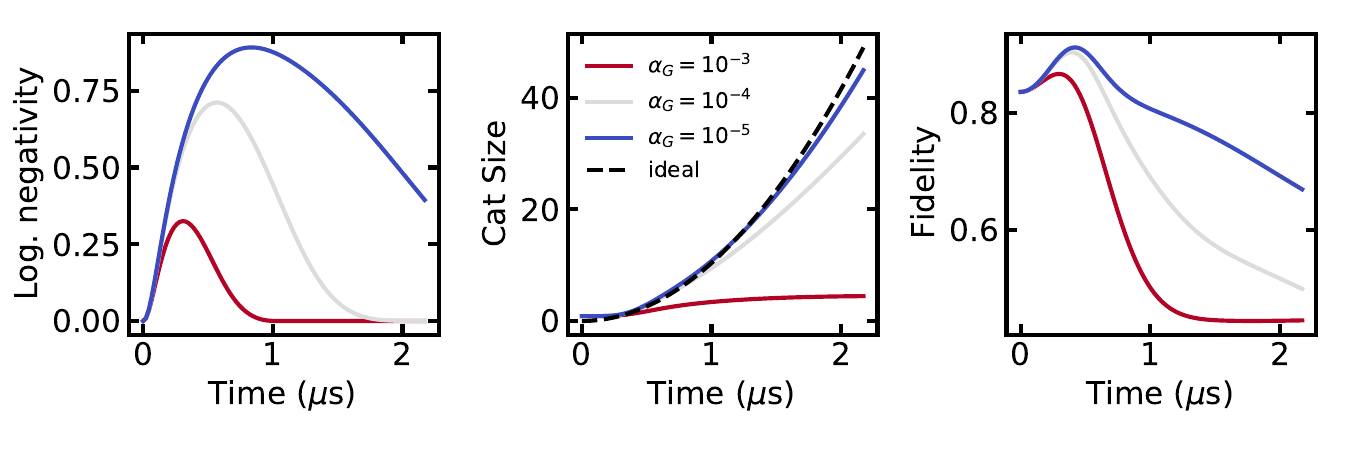}
  \end{center}
  \caption{
  { Same plots as in Figs.~3(a-c) of the main text for an initially thermally excited magnon state with $n_\mathrm{th}\simeq0.1$ (top row) and $n_\mathrm{th}\simeq0.4$ (bottom row) corresponding to temperatures of 10~mK and 20~mK, respectively.} 
  }
  \label{fig:TemperatureDependence}
\end{figure}
Fig.~\ref{fig:TemperatureDependence} benchmarks the qubit-magnon entanglement and cat-state fidelity at higher temperatures than adopted in Fig.~\ref{fig:CatStates} of the main text.
A temperature of 10~mK (20~mK) corresponds to an initially thermally populated magnon state with $n_\mathrm{th}\simeq 0.1\ ( 0.4)$.
We find that large cat states of size $S\sim10-40$ may still be formed with a fidelity above $80 \%$ for typical Gilbert damping parameters $\alpha_{G}\sim10^{-5}-10^{-4}$~\cite{tabuchi2015coherent,klingler2017gilbert,rameshti2021cavity}.

\end{document}